**Zoltán Bartha**

# Configurations, not thresholds: the middle-income trap in the CEE members of the OECD

Abstract

The middle-income trap is usually identified by comparing a country's income to a threshold. This reveals that convergence has stalled but not where the blockage lies, and it leaves unexplained why the trap is still debated in countries that have already passed the threshold. This study treats the trap as a configuration problem, measuring development along three dimensions – future, outside, and inside potential – with indices built from standardised indicators. The indices are calculated for 34 OECD members in 2000, 2010, and 2020, so that changes in configuration become visible, and the resulting country clusters are interpreted through factor analysis. Two traps and two transition strategies emerge. The first strategy attracts foreign investment and technology, but can itself become a trap: domestic value added stays limited and rising wages erode export competitiveness. The second adds a developmental state focused on the quality of education, building the human capital that raises productivity and innovation. All six CEE members of the OECD had adopted the first strategy by 2010; by 2020 only Estonia had completed the second. Configuration levels and changes track relative income convergence.

## 1. Introduction

Garrett (2004) suggested that there are two ways to grow in the global economy: one can either gain a competitive edge by relying on unique skills, and capabilities or compete through low costs, especially low wages. Those who are caught in the middle, are in danger of being left behind. A few years later Gill & Kharas (2007) introduced the concept of the middle-income trap (MIT), describing the economic stagnation observed in certain economies that, after a period of rapid growth, experience a significant deceleration upon reaching middle-income status. The slowdown is generally viewed as a transition problem: as growth leads to rising wages, the growth strategy based on low wages is no longer sustainable and the policy challenge is to design a new approach. The concept was tested by a series of empirical studies, which typically used per capita national income as the measure of economic development. The threshold between the upper-middle income and the high income group can be drawn in absolute terms (e.g., Aiyar et al., 2013; Felipe et al., 2012) or in relative terms (usually compared to the US, e.g., Bulman et al., 2017). The World Bank adopted the Atlas method (World Bank, 2024a) for the classification of countries. According to them, the high income threshold stood at USD14,005 GNI per capita in 2025 (World Bank, 2024b).

Although all 11 former socialist EU members had surpassed the World Bank's high-income threshold by 2025, overcoming the middle-income trap (MIT) remains a significant topic of public debate in these countries (Kouli & Müller, 2023). This apparent contradiction can be explained by two key findings. First, some empirical studies found evidence of two different MITs at two different income levels (Eichengreen et al., 2013; Hannum, 2024), with one situated near the current income level of Central and Eastern European (CEE) countries. Second, the most recent World Bank report presents the 3i model of development (first invest, then infuse, and finally innovate), which suggests that developing countries require at least two strategic transformations to achieve high-income status.

This study treats the middle-income trap as a configuration problem: while an income threshold shows that convergence has stalled, this approach helps reveal where in an economy's structure the blockage lies. Developing this idea, the study makes three contributions to the MIT literature. The first is about measurement. The FOI model (Bartha, 2024) tracks development along three dimensions – future, outside, and inside potential – instead of the single income figure the MIT literature usually relies on. The model is applied here at three observation points – 2000, 2010, and 2020 – so that an economy's trajectory over two decades becomes visible, not just its position at a single moment. It also helps explain why the MIT is still debated in countries that have already passed the income threshold. The second contribution is theoretical. The main explanations of the trap disagree about its causes but share a common structure. Each describes a growth strategy that works at middle income, a different strategy needed to move beyond it, and a country that fails to switch from one to the other. That shared structure is what makes the configuration reading possible: the paper sets it out and links each explanation to one of the three dimensions. The third contribution is empirical. The analysis covers 34 OECD economies and focuses on the six former socialist members among them: Czechia, Estonia, Hungary, Poland, Slovakia, and Slovenia. It identifies the strategies that separate the economies that escaped from those that did not and checks the result against income data.

The remainder of this study is structured as follows. Section 2 reviews the MIT literature: the competing definitions, the difficulties they raise, and the theoretical explanations offered for the phenomenon. then sets out the theoretical framework. Section 3 describes the indices, how they are built, and the sample. Section 4 presents the results: the configurations of the OECD economies and how they change, the factors that distinguish the strategies behind them, the position of the CEE members, a check of the configurations against income data, and the robustness of the classification. Section 5 concludes.

## 2. Literature review

The MIT occurs when fast-growing developing economies slow down as they reach the middle-income stage of development (Gill & Kharas, 2007). When the trap is described, there are at least two technical elements and one theoretical aspect that need to be defined. One obvious technical question is related to the stages of development: what rule we use to categorise countries into low-, middle-, and high-income groups. The other technical question is about the pace of growth (or the amount of time spent in one development stage): what growth rate is needed for an economy to avoid being stuck at the middle-income stage. The theoretical challenge is to explain why the MIT occurs and how it can be overcome.

Per capita income is the standard metric for classifying stages of economic development. This income can be measured using GDP at purchasing power parity (PPP) or GNI. The World Bank provides GNI data in USD, calculated using the Atlas method, which employs a three-year moving average of exchange rates. Two main approaches exist for defining income-based development stages: absolute and relative. The absolute approach establishes fixed income thresholds (Aiyar et al., 2013; Eichengreen et al., 2013; Felipe et al., 2012; Jayadevan et al., 2024). For example, Felipe et al. (2012) used GDP per capita in 1990 PPP USD, defining middle-income countries as those with per capita GDP between $2,000 and $11,750. The relative approach, conversely, compares a country's per capita income to a benchmark, such as the US (Arias & Wen, 2016; Bulman et al., 2017) or the world average (Gao et al., 2024). Regardless of whether an absolute or relative approach is used, the field commonly distinguishes between lower-middle and upper-middle income groups. The absolute approach is more effective for analysing the causes of economic slowdowns, while the relative approach is more suitable for convergence analysis (Im & Rosenblatt, 2015).

If GDP per capita grows by 5 percent on a yearly average, it takes almost 40 years to go from $2,000 to $11,750. This implies that identifying a development trap requires a long-term perspective. Research indicates that recognizing such a trap may take 40-50 years (Glawe & Wagner, 2016). The process may be accelerated if we focus on the growth rates instead of the time spent at a given development stage. Growth slowdowns can be detected as sudden and sustained deviation from a trend. This trend can be established using historical growth data (Eichengreen et al., 2013) or model-based calculations (Aiyar et al., 2013). If these deviations persist for 5-7 years, they may signal a development trap.

Several theories attempt to explain the MIT. When introducing the concept, Gill & Kharas (2007) clearly state that different growth strategies are required at different development stages. The need for a policy transition is often linked to the competitive advantage of different nations. Initially, low wages provide a competitive edge for less developed economies, particularly in mature, highly standardized industries. Growth in this interpretation slows when the low-wage advantage is lost (Gill & Kharas, 2007; Győrffy, 2022; Kang & Paus, 2020). One common explanation for rising wages is the Lewis-effect: growth is generated when surplus labour is reallocated to the more efficient sectors, but this surplus eventually diminishes, exhausting the possibility of further growth without wage rises (Ekanayake, 2021; Lewis, 1954). The Lewis-effect is connected to the concept of the Kuznets curve (Kuznets, 1955; Ozturk, 2016). Kuznets suggests that the initial development phase – characterised by labour moving towards more efficient sectors, capital becoming concentrated, and skill gaps developing – leads to higher inequality. However, if the economy can reach a second transition phase, inequalities typically decrease. This is usually due to redistributive policies and the accumulation of human capital (Hu et al., 2023).

Beyond the point where wages rise, growth slows unless total factor productivity (TFP) increases (Lee & Park, 2024). One classical explanation for why TFP growth varies across countries is Tamura's model, which attributes unequal global income growth to differences in economies' ability to adopt and diffuse technology (Lucas, 2000; Tamura, 1996). The World Bank's report on the MIT (2024c) introduces the 3i model of economic transition. This model hypothesises that initial growth is primarily driven by capital investments. In a second stage, investments need to be coupled with increased infusion of technology. Finally, in the third stage, innovation capability becomes essential for continued growth. This theoretical model aligns with some empirical findings suggesting the existence of not one, but two distinct MITs (Eichengreen et al., 2013; Hannum, 2024). The 3i model is based on the Schumpeterian hypothesis, which argues that sustained growth, and TFP increases result from more efficient firms replacing less efficient ones. Aghion & Howitt (1992) introduced the Schumpeterian hypothesis into growth theory and Akcigit & Kerr (2018) later expanded it to include the impact of efficient incumbent firms. A comprehensive survey of these arguments is provided by Agénor (2017), who organises the causes of middle-income traps into seven categories – diminishing returns to capital, exhaustion of cheap labour and imitation gains, insufficient quality of human capital (Redding, 1996), distorted incentives and misallocation of talent (Agénor & Canuto, 2015), inadequate contract enforcement, lack of advanced infrastructure, lack of access to innovation finance – and shows that they are complementary rather than competing explanations.

There are two main branches of theory that attempt to explain why some countries transition between growth stages more successfully than others: institutional theory and world order theory. Both are related to the problem of incumbency: the tendency of powerful established groups to manipulate outcomes in their favour, ultimately hindering efficient growth. Institutional theory focuses on a country's internal norms and rules arguing that incumbents can have both the political and economic means to obstruct transition (Acemoglu, Aghion, et al., 2006; Mickiewicz, 2023; Song et al., 2023). They protect their financial interests tied to existing, less efficient firms and sectors by implementing protectionist regulations or rules that reduce competition. Elites are more successful at slowing down the transition when they have a stronger grip on the political system (Acemoglu et al., 2019). Powerful rent-seeking groups can even extract resources from the economy which can further slow growth if the rents are transferred abroad (Ang & Dong, 2023).

World order theory suggests that countries have unequal power to pursue their interests on the global stage. This theory suggests a global hierarchy, with more developed countries at the top. This hierarchical structure leads to an unequal distribution of profits, hindering growth in developing countries (Gao et al., 2024). Anastasi (2023) reframes the MIT as a "semi-periphery trap". The core issue of the semi-periphery trap is that middle-income countries integrated into the global economy cannot retain enough surplus generated inside the country. As a result, they lack the resources needed to invest in the capabilities required for achieving high-income status.

In summary, the theoretical explanations for the MIT are highly complex, encompassing societal structures, regulatory mechanisms, and hierarchical relationships between countries. However, the measurement of the MIT is relatively simple, typically relying on variations of per capita income to determine whether a country is trapped. While the methodology for calculating gross income is sophisticated, national income figures alone cannot capture the full complexity of the theoretical constructs associated with the MIT. This discrepancy explains the ongoing public debate in countries that have already achieved high-income status. Therefore, developing a more comprehensive measurement approach could significantly improve our understanding of the MIT.

### 2.1. Theoretical framework

The middle-income trap is, at its core, a claim about dynamics. Whether defined in absolute terms, in relative income terms, or through the incidence of sustained growth slowdowns, the trap is identified from the trajectory of an economy: a period of successful catch-up followed by a prolonged stall before the high-income frontier is reached. Trajectory-based identification is the natural empirical strategy, and it has produced the canonical lists of trapped and escaped economies. It also has two well-documented limitations (Agénor, 2017; Glawe & Wagner, 2016). First, its verdicts are sensitive to threshold and period choices: small changes in the income cut-offs or the slowdown criteria reclassify a substantial number of countries. Second, it is diagnostic in nature. A stalled trajectory reveals that convergence has stopped, but does not indicate where in the economy's structure the blockage lies. The escape from the trap is never a resumption of the old growth model at a higher speed – every theoretical account of the trap agrees that it requires a qualitative reorganisation of how growth is generated. Detecting that reorganisation, or its absence, requires looking beneath the income trajectory at the structure that produces it.

The theoretical literature on the trap is heterogeneous in mechanism but uniform in architecture. Gill and Kharas (2007) describe middle-income economies as squeezed between two viable configurations – low-wage exporters below them and high-innovation economies above. The World Bank's (2024c) 3i framework makes the stage logic explicit: economies must move from investment-driven growth through the infusion of foreign technology and know-how to domestic innovation and the trap is a failed transition between stages. The Lewis-effect argument reaches the same structure from the labour market: once surplus labour is exhausted and wages rise past the turning point (Lewis, 1954), the cost-based, externally financed configuration loses viability by construction, and continued development requires capabilities that the old configuration did not need and did not build. This switch in capabilities requires better financing for higher-value added businesses (Agénor, 2017), improved quality of human capital (Redding, 1996), and mechanisms that redirect talent toward higher value-added activities (Agénor & Canuto, 2015).

The Schumpeterian growth theory formalises the switch. Acemoglu et al. (2006) show that the institutions and policies appropriate to investment-based catch-up growth differ from – and can actively obstruct – those required for innovation-based growth near the frontier, so that economies which fail to reorganise become trapped in non-convergence (see also Aghion & Bircan, 2017). The institutional literature makes the same point about the state and the coordination system: the institutional configuration that successfully attracts external capital and delivers macroeconomic stability during catch-up is not the configuration that sustains frontier development. The latter relies on education quality, the rule of law, and the efficiency of domestic resource allocation (Győrffy, 2022; Kang & Paus, 2020). Finally, the world-systems and dependent-development tradition (Wallerstein, 1974) frames the semi-periphery as a structural position in the international division of labour: an economy integrated into world markets on the basis of externally owned capital and cost advantages faces a ceiling that is a property of its configuration.

These six accounts converge on the proposition that the middle-income trap is a configuration problem. Each theory identifies a growth configuration that is viable for entering middle income, identifies the different configuration required for leaving it, and locates the trap in the failure to move from one to the other. Where the theories differ is in which capability they treat as the binding constraint of the transition – world-market position, technological absorption, labour costs, quality of human capital, innovation capacity, or institutional quality. This gap can be filled by a structural capability framework. The Future, Outside, Inside (FOI) model adopted here serves as a measurement surface on which the configurations that the trap theories reason about become observable and comparable. Its three dimensions correspond to the three areas where the theories place their binding constraints. The Outside potential (O) measures an economy's world-market position and therefore operationalises the external, dependent-development side of the trap arguments: the 2i infusion stage, the externally financed catch-up configuration, the semi-peripheral position. The Future potential (F) measures the preconditions of innovation-based growth and therefore operationalises the frontier configuration that Schumpeterian and 3i accounts identify as the destination of a successful transition. The Inside potential (I) measures the quality of domestic resource allocation and welfare and therefore operationalises the institutional and coordination arguments about what must hold domestically for either configuration to function.

Anticipating the measurement strategy set out in Section 3, a country's configuration is denoted by the three letters F, O, I, capitalised where the corresponding potential is high and lower-cased where it is low. So fOi denotes low future and inside potential with high outside potential. The eight possible combinations form the configuration space in which development is tracked. Table 1 summarises the mapping.

Table 1. Middle-income trap mechanisms and their representation in the FOI framework

| Theoretical account | Trap mechanism | Binding constraint of the transition | FOI representation |
|---|---|---|---|
| Gill & Kharas (2007) | Squeezed between low-wage and high-innovation configurations | Switch from scale/cost to specialisation/innovation | High O without F → trap; escape requires building F |
| World Bank (2024c) 3i | Failed transition between investment → infusion → innovation stages | Absorptive, then innovative capability | O captures infusion (2i); F captures innovation (3i); I captures the investment climate (1i) |
| Lewis-effect (Lewis, 1954) | Surplus labour exhausted; wage growth erodes cost advantage | Viability of the cost-based external configuration | fOi configuration is transitory by construction; wage pressure undermines O unless F is built |
| Schumpeterian growth (Agénor, 2017; Acemoglu et al., 2006) | Investment-based institutions obstruct innovation-based growth | Reorganisation of institutions and firm selection, insufficient quality of human capital | fOi→FOi transition is the investment→innovation switch made observable |
| Institutional accounts (Agénor, 2017, Kang & Paus, 2020; Győrffy, 2022) | Catch-up institutions ≠ frontier institutions | Governance quality, education, coordination, distorted incentives and misallocation of talent, lack of access to innovation finance | I measures allocation quality; F measures long-horizon institutional investment |
| World-systems / dependency | Semi-peripheral position in the | Ownership and control of the growth engine | fOi operationalises externally driven, dependent development |
| | international division of labour | | |

Source: own work

Three boundaries of this approach should be stated at the outset. First, the FOI framework complements income-based trap identification, it does not replace it. A country's configuration says little about its income level by construction and this independence is analytically useful, because it allows the framework to register the cases where income status and structural capability diverge (such as high-income economies whose configurations have decayed to those of trapped middle-income countries).

Second, the FOI indices are relative measures, so a country's configuration expresses its position among the economies it is compared with, not an absolute level. This construction makes the reference group a substantive choice. Configurations are comparable only when the group is reasonably homogeneous in institutional setting and in the external conditions under which its members develop, because it is against that common backdrop that a high or low potential acquires meaning. It is therefore essential to hold the reference group to the OECD membership, a set of economies that develop under broadly common rules of trade, capital mobility and economic governance, and for which comparable indicators are available across the whole period. This also fixes the group over time, so that movement between configurations reflects a country's own trajectory rather than a change in the comparison set. Extending the analysis to economies developing under materially different external and institutional conditions would shift the standardisation benchmark and make cross-group configurations difficult to interpret.

Third, the configuration lens is descriptive rather than causal: observing that escaping countries share a configuration change does not identify the direction of causation between capability building and growth. Section 4.4 therefore examines whether configurations and configuration changes co-move with relative income levels and convergence outcomes in the expected directions, and it acts as a consistency check on the mapping proposed here.

## 3. Data and methods

Before describing how the three indices are built, it is worth stating why three observations, spaced a decade apart, are an adequate basis for studying a process usually understood as dynamic. A trajectory is a continuous object and demands long, high-frequency series; a configuration is a discrete object and can be observed at intervals. The FOI framework deliberately mixes components with different speeds of change – norms that move over decades (work culture, social responsibility), institutions that move over years (education and health systems, research arrangements), and policies that can move quickly (sectoral expenditures) – so that the composite indices themselves are slow-moving by construction. Decadal observation therefore samples the phenomenon at approximately its natural frequency. Measuring a country's configuration in 2000, 2010 and 2020 captures the state of a process whose meaningful changes unfold over such horizons. The dynamics of development then appear in the data as movement – or persistence – in the configuration space.

Three analytically distinct patterns become observable. Persistence in a low configuration across two decades (a country that has low potential across all three pillars at all three observation points) is the configurational signature of the trap, translating the duration criterion of the income-based definitions into structural terms. Transition between configurations (from a low potential in 2000, to a high one by 2020) is the observable form of the strategy switch that all six theoretical accounts identify as the escape. And regression across configurations – which, as Section 4 shows, is empirically common – is the observable form of the erosion mechanisms (Lewis-type wage pressure, relative decline against a moving frontier) that the theories predict for economies that fail to reconfigure. In this reading, the eight-cell FOI space functions as a coarse state space for the development process, and the country transitions between 2000, 2010 and 2020 constitute its observable dynamics.

### 3.1 Constructing the FOI indices

The FOI model resolves development into three potentials: Future (F), the capacity for innovation-based growth; Outside (O), the economy's position in the world market; and Inside (I), the quality of domestic resource allocation and welfare. Each potential is measured by an index built from a set of indicators through a transparent, fixed procedure that has been applied in earlier FOI studies (Bartha, 2024; Bartha & S. Gubik, 2014), which makes results comparable across those studies and across time.

The indicators were selected in two stages. In the first, conceptual stage, the candidate development factors were drawn from a review of the development-economics literature, attending to supply- and demand-side determinants of competitiveness and to institutional influences. This literature-based list was refined through a structured expert elicitation: an external panel first assembled, by brainstorming, a set of fifty candidate indicators judged to plausibly measure the underlying factors while being available for many countries and periods. These fifty indicators were then rated by twenty-eight experts on a seven-point importance scale, separately for each of the three potentials, according to how decisive each factor was for an economy's future competitiveness (F), current world-market position (O), and current welfare (I). Each indicator was assigned to the single potential for which it received the highest mean rating, and only indicators scoring above the overall mean were retained, so that the indices are built exclusively from factors the experts judged to be of high importance. In the second, operational stage, each retained factor was matched to a concrete measure using explicit criteria: cross-country coverage of the sample, source credibility and methodological continuity, and availability at or near each observation year.

Each indicator is standardised to a one-to-seven scale using the min-max method: the lowest observed value across the sample is set to 1, the highest to 7, and the remainder are placed proportionately between. Indicators for which a lower raw value is preferable (for example, industrial disputes, the ecological footprint, or exchange-rate variance) are inverted before scaling, so that a higher standardised score always denotes better performance. Each index is the unweighted arithmetic mean of its standardised components. The equal-weight, arithmetic-mean form is the established FOI convention and treats the components of a potential as partial substitutes; the robustness of the resulting classification to this choice, and to the min-max anchoring, is examined in Section 4.5.

Because some indicators were discontinued or unavailable in earlier years, the exact component set differs slightly across the three observation points. The composition is reported in full in Table 2. Where a component is missing for a particular country and year, the index is the mean of the available components. These differences reflect data availability rather than a change in the model, and Section 4.5 confirms that they do not materially affect the classification. The indicators are drawn from several established international databases; the primary source of each component is given in Table 2, and the full indicator-to-source correspondence in Appendix 1.

Table 2. Composition of the FOI indices, 2000–2020

| Potential | Component | 2000 | 2010 | 2020 | Primary source |
|---|---|---|---|---|---|
| F | Social responsibility | ● | ● | ●● | IMD-WCY |
| F | Work ethic / culture (↓) | ● | ● | ● | IMD-WCY |
| F | Energy efficiency | ● | ● | ● | IMD-WCY |
| F | Education expenditure per capita | ● | ● | ● | OECD |
| F | Ageing of society | | | ● | IMD-WCY |
| F | Renewable energy | | | ● | IMD-WCY |
| F | Healthy life expectancy | ● | ● | ●● | WHO |
| F | Ecological footprint (↓) | ● | ● | ● | Global Footprint Network |
| F | R&D expenditure per capita | ● | ● | ● | OECD |
| F | R&D potential (researchers, patents) | ● | ● | ● | IMD-WCY / OECD |
| F | Efficiency of the education system | ● | ● | ● | OECD-PISA / IMD-WCY |
| O | Trade openness | ● | ● | ● | OECD / World Bank |
| O | Country credit rating | ● | ● | ● | Institutional Investor |
| O | Financial sector stability | ● | ● | ● | IMD-WCY / WEF-GCI |
| O | Exchange-rate stability (↓) | ● | ● | ● | IMF (IFS) |
| O | Language skills (TOEFL) | | ● | ● | ETS |
| I | Government efficiency | ● | ● | ●● | IMD-WCY / WEF-GCI |
| I | Social wellbeing / quality of life | ● | ● | ● | IMD-WCY |
| I | Tax burden | ● | ● | ● | IMF / IMD-WCY |
| I | Pension system sustainability | | ● | ● | IMD-WCY |
| I | Level of development (GDP PPP p.c.) | ● | ● | ● | IMF (WEO) |
| I | Real GDP growth | ● | ● | ● | IMF (WEO) |
| I | Availability of capital | ● | ● | ● | IMD-WCY / WEF-GCI |
| I | Labour-market flexibility (↓ rigidity) | | ● | ● | World Bank / IMD-WCY |
| I | Employment / labour force | ● | ● | ● | OECD (LFS) |
| I | Skilled labour availability | ● | ● | ● | IMD-WCY |
| | Number of components (F / O / I) | 9/4/8 | 9/5/10 | 13/5/11 | |

*Notes: ●● denotes a factor represented by two indicators in that year, each entering the index mean separately. (↓) denotes a reverse-coded indicator, for which a lower raw value is preferable and which is inverted before standardisation. Where a component is unavailable for a given country and year, the index is computed as the mean of the available components.*

Source: own compilation

## 3.2 Sample and analytical strategy

The sample comprises the 34 economies that were OECD members in 2010. This fixes the country set across the whole period and avoids composition changes from later accessions (Colombia, Costa Rica, Latvia, Lithuania), for which 2000 data are largely unavailable. The Central and Eastern European (CEE) subgroup therefore is constrained to six former socialist countries: Czechia, Estonia, Hungary, Poland, Slovakia, and Slovenia. The OECD is a club of developed countries, but many of its members have been close to the high income threshold published by the World Bank. In 2000 the threshold stood at $9,265 and eight countries (Chile, Czechia, Estonia, Hungary, Mexico, Poland, Slovakia, Turkey) were below this level. By 2010 the threshold moved to $12,195, but only three countries lay below it (Chile, Mexico and Turkey). Finally, by 2020 only Mexico and Turkey were below the $12,695 high income threshold. The advantage of focusing on the OECD is that it provides a better understanding of the development choices countries near the threshold can make. Another advantage is that the paths taken by more developed OECD members can serve as examples for emerging economies.

Table 3. The Future-, Outside-, and Inside-index of the 34 evaluated countries (2000-2020)

| Country | F2020 | F2010 | F2000 | O2020 | O2010 | O2000 | I2020 | I2010 | I2000 |
|---|---|---|---|---|---|---|---|---|---|
| Australia | 3.8 | 4.6 | 4.5 | 5.3 | 5.3 | 4.6 | 4.6 | 4.4 | 4.3 |
| Austria | 4.4 | 5.1 | 5.3 | 5.1 | 5.4 | 4.2 | 3.9 | 4 | 4.7 |
| Belgium | 3.8 | 4.2 | 5.1 | 4.9 | 5.6 | 4.9 | 3.6 | 3.5 | 4.3 |
| Canada | 4 | 4.2 | 4.9 | 4.9 | 5.4 | 5 | 4.6 | 4.5 | 4.7 |
| Chile | 3.6 | 3.8 | 3.9 | 3.9 | 5 | 4 | 3.8 | 4.1 | 2.9 |
| Czechia | 3.8 | 3.4 | 3.1 | 4.2 | 5 | 2.4 | 3.2 | 3.6 | 3.3 |
| Denmark | 4.9 | 5.3 | 5.2 | 5 | 5.8 | 4.4 | 4.7 | 4.3 | 4.8 |
| Estonia | 4.2 | 3.2 | 3.1 | 4.7 | 4.9 | 3.7 | 3.6 | 3.1 | 3.3 |
| Finland | 4.6 | 5.4 | 5.6 | 5.1 | 5.7 | 4.6 | 4.9 | 4 | 5.1 |
| France | 4.2 | 4.7 | 5 | 4.3 | 4.5 | 4 | 3.5 | 3 | 4.3 |
| Germany | 4.4 | 4.8 | 4.9 | 4.7 | 5.3 | 4.3 | 4.5 | 3.7 | 4.3 |
| Greece | 3.3 | 3.1 | 3 | 2.9 | 3.7 | 2.8 | 1.9 | 2.5 | 3.2 |
| Hungary | 3.1 | 3.2 | 3.4 | 4.4 | 4.6 | 3.2 | 2.6 | 2.5 | 3.4 |
| Iceland | 5.3 | 5.8 | 5.6 | 4.2 | 2.3 | 4.1 | 5 | 4.4 | 5.1 |
| Ireland | 4.3 | 4.2 | 4.1 | 4.6 | 4.2 | 4.7 | 5 | 3.9 | 4.5 |
| Israel | 4.5 | 3.6 | 4.2 | 4.6 | 4.9 | 4.1 | 4.1 | 4.1 | 4.3 |
| Italy | 3.5 | 3.7 | 3.9 | 3.5 | 3.8 | 3.2 | 2.7 | 2.7 | 3.6 |
| Japan | 4.7 | 5.5 | 5.6 | 3.7 | 3.7 | 3.5 | 4.1 | 4 | 3.5 |
| Korea | 4.3 | 4.5 | 4 | 4.3 | 4.3 | 3.5 | 3.8 | 3.3 | 3.3 |
| Luxembourg | 3.8 | 6.1 | 5.4 | 6.1 | 6.6 | 5.8 | 4.6 | 4.5 | 5.7 |
| Mexico | 3 | 2.6 | 3 | 4.1 | 4 | 3 | 3.3 | 2.9 | 2.4 |
| Netherlands | 4.3 | 4.9 | 5.1 | 5.3 | 5.5 | 5 | 5.3 | 3.8 | 4.6 |
| New Zealand | 4.5 | 4.4 | 4.7 | 5.1 | 4.5 | 4.5 | 4.8 | 4 | 4.1 |
| Norway | 4.7 | 5.5 | 5.2 | 4.9 | 5.7 | 5 | 4.9 | 4.1 | 4.6 |
| Poland | 3.7 | 3.1 | 3.2 | 4 | 4.4 | 3.2 | 3.1 | 3.1 | 2.8 |
| Portugal | 3.9 | 3.7 | 3.6 | 3.7 | 4.3 | 3.9 | 3.1 | 2.9 | 3.4 |
| Slovakia | 3.4 | 3.3 | 3.6 | 4.8 | 4.8 | 2.6 | 2.9 | 3.3 | 3.1 |
| Slovenia | 4 | 3.7 | 4.1 | 4.5 | 5.1 | 3.2 | 3.2 | 2.7 | 3.3 |
| Spain | 3.2 | 3.7 | 3.7 | 4 | 4.2 | 3.7 | 3.1 | 3 | 4 |
| Sweden | 4.9 | 5.5 | 5.6 | 4.9 | 5.2 | 4.8 | 4.6 | 4.1 | 4.7 |
| Switzerland | 5.2 | 5.9 | 5.9 | 5.4 | 5.4 | 4.8 | 5.6 | 4.9 | 4.9 |
| Turkey | 3.1 | 3 | 2.9 | 3.2 | 3.6 | 1.9 | 3.1 | 3.1 | 2.6 |
| United Kingdom | 3.8 | 4.3 | 4.8 | 5.3 | 4.3 | 5 | 4.7 | 3.6 | 4.1 |
| United States | 3.9 | 4.1 | 5 | 5.4 | 4.3 | 5 | 5.3 | 4.5 | 4.5 |

Source: own calculations, extending Bartha (2024)

This study employs the FOI indices for two primary purposes. First, clusters are set up to describe potential growth strategies chosen by OECD members. Due to the standardization procedure, the indices are scaled from 1 (minimum) to 7 (maximum). By applying an interval halving technique, countries are categorized into eight distinct clusters based on whether their FOI indices have low (less than 4) or high (4 or above) values. The cluster with the lowest development level is characterized by low values across all three indices. There are three clusters with one high index, three with two high indices, and one cluster, representing the highest development level, with high values across all three dimensions. Despite its lack of methodological sophistication, the interval-halving method produces relatively stable clusters, delivering results broadly consistent with data-driven clustering methods at the coarse level (se Section 4.5, and Appendix 3).

Second, factor analysis is conducted to identify the underlying factors that explain shifts between these clusters. These factors, which explain the likelihood of an index having a low or high value, help in describing and interpreting the strategic choices available for countries seeking to overcome the MIT.

## 4. Results and discussion

The results are presented in five sections. The first depicts the evolution of FOI clusters within the OECD from 2000 to 2010 and then to 2020, establishing a baseline for the development strategies adopted across the sample. The second uses factor analysis to provide a more nuanced understanding of what distinguishes those strategies. The third assesses the performance of the CEE OECD members against them. The fourth examines whether the configurations and their changes co-move with relative income levels and convergence, as a consistency check on the mapping set out in Section 2.1. The fifth reports the robustness of the classification to the threshold, to the standardisation, and to measurement error.

### 4.1. Cluster formation and evolution in the OECD

Using the FOI indices from Table 3, the thirty-four OECD economies were assigned to clusters by the interval-halving rule, an index counting as high when it reaches four. The full classification for the three observation years is shown in Table 4. The configuration space is populated more richly and unevenly than a single high–low split would suggest.

Table 4. The FOI clusters of the OECD, 2000–2020

| Cluster | 2000 | 2010 | 2020 |
|---|---|---|---|
| FOI (all high) | AUS, AUT, BEL, CAN, CHE, DEU, DNK, FIN, FRA, GBR, IRL, ISL, | AUS, AUT, CAN, CHE, DNK, FIN, LUX, NOR, NZL, SWE, USA *(n = 11)* | CAN, CHE, DEU, DNK, FIN, IRL, ISL, ISR, NLD, NOR, NZL, SWE *(n = 12)* |
| | ISR, LUX, NLD, NOR, NZL, SWE, USA *(n = 19)* | | |
| FOi (F, O high) | — | BEL, DEU, FRA, GBR, IRL, KOR, NLD *(n = 7)* | AUT, EST, FRA, KOR *(n = 4)* |
| FoI (F, I high) | — | ISL, JPN *(n = 2)* | JPN *(n = 1)* |
| fOI (O, I high) | — | CHL, ISR *(n = 2)* | AUS, GBR, LUX, USA *(n = 4)* |
| Foi (F high) | JPN, SVN *(n = 2)* | — | — |
| fOi (O high) | CHL *(n = 1)* | CZE, ESP, EST, HUN, POL, PRT, SVK, SVN *(n = 8)* | BEL, CZE, ESP, HUN, MEX, SVK, SVN* *(n = 7)* |
| foI (I high) | ESP *(n = 1)* | — | — |
| foi (all low) | CZE, EST, GRC, HUN, ITA, KOR, MEX, POL, PRT, SVK, TUR *(n = 11)* | GRC, ITA, MEX, TUR *(n = 4)* | CHL, GRC, ITA, POL*, PRT, TUR *(n = 6)* |

*Notes: Clusters are named by the three potentials (F, O, I), capitalised where the potential is high (index ≥ 4) and lower-cased where low. A dash denotes an empty cluster. The * marker denotes a boundary case, where the decisive index lies within 0.1 of the threshold: Poland's 2020 O-index (4.00 displayed, 3.9995 unrounded) falls just below it, and Slovenia's 2020 F-index (3.99) likewise.*

*Country codes: AUS Australia, AUT Austria, BEL Belgium, CAN Canada, CHE Switzerland, CHL Chile, CZE Czechia, DEU Germany, DNK Denmark, ESP Spain, EST Estonia, FIN Finland, FRA France, GBR United Kingdom, GRC Greece, HUN Hungary, IRL Ireland, ISL Iceland, ISR Israel, ITA Italy, JPN Japan, KOR Korea, LUX Luxembourg, MEX Mexico, NLD Netherlands, NOR Norway, NZL New Zealand, POL Poland, PRT Portugal, SVK Slovakia, SVN Slovenia, SWE Sweden, TUR Turkey, USA United States.*

Source: own work

In 2000 the distribution is bimodal. Nineteen economies occupy the all-high FOI cluster and eleven the all-low foi cluster, but four sit in intermediate positions. The eleven foi economies include all eight OECD members then below the World Bank's high-income threshold of $9,265 per capita, joined by three Mediterranean economies (Italy, Greece, Portugal) and Korea. In the FOI framework this all-low placement signals exposure to the middle-income trap, and

persistence in it over the full period is the configurational trap signature. Section 4.4 confirms that the economies which remained there also stagnated or fell in relative income.

By 2010 the intermediate clusters fill in. Two new configurations become prominent: eight economies hold high outside potential alone (fOi) and seven hold high future and outside potential together (FOi). The fOi cluster is the more consequential for this study, since it absorbs five of the six CEE members, which reached it by raising their O-index through deeper trade and financial integration. This outside-oriented shift corresponds to the investment-attraction and export-led strategies of the trap literature (Gill & Kharas, 2007; World Bank, 2024c), strategies that the same literature warns are insufficient on their own: the low-cost advantage erodes as wages rise, and heavy export reliance exposes an economy to the volatility of its markets (Mellár, 2001). Whether the CEE economies converted this external opening into the broader capability base needed to escape the trap is the question Section 4.3 takes up.

Several high-income economies occupy the fOI cluster, combining strong outside and inside potential with a future index below the threshold. The future index captures a broad set of structural foundations for long-run development (human capital breadth, environmental sustainability, and the social underpinnings of growth) whose composition is discussed by Bartha (2024). This cluster lies outside the CEE development paths that concern this study.

Movement in the FOI space runs in both directions. Two economies illustrate the escape route the trap theories describe (Table 1): Korea reached the FOi cluster by 2010 and Estonia by 2020, each combining the external opening with a rising future potential – a strategy switch from outside-led catch-up to innovation-oriented growth made observable as a cluster transition. But regression is equally present. Belgium slipped from FOI in 2000 to FOi in 2010 and fOi by 2020 as its inside and then future potentials fell below the threshold, while its world-market position held. This gradual erosion of structural capability beneath an unchanged high-income surface is the divergence between income and configuration that motivates the framework.

Two countries sit very close to the threshold. Poland's 2020 outside index is 4.00 to two decimals but 3.9995 unrounded, so under the strict rule it falls just inside the all-low foi cluster rather than the fOi cluster its CEE peers occupy. A Monte-Carlo perturbation of its components (Section 4.5) assigns it to foi and fOi with almost equal probability (51/49), confirming that this is a genuine boundary rather than a clear reversal. Slovenia's 2020 future index, at 3.99, places it in fOi, but the same perturbation gives it a 39 per cent probability of reaching the future-high FOi cluster, so its position is a cusp rather than a settled low. Treating these two as boundary cases rather than forcing them into a single bin is both more accurate and more informative than the threshold alone would allow.

## 4.2. Factor analysis and movements between configurations

The fOi and FOi clusters represent potential transition strategies toward high income status. To explore the characteristics of these strategies, I examined the factors influencing the FOI indices. A principal component analysis (PCA) with Varimax rotation and Kaiser normalization was performed for this purpose, and the key findings are presented in Table 5.

Table 5. Influencing factors of the Future, Outside and Inside pillars of development

| Factors of the F-index | Factors of the O-index | Factors of the I-index |
|---|---|---|
| **F1 Human capital**<br>Labour productivity (GDP PPP/employed person/hour)<br>Overall productivity (GDP PPP/employed person)<br>Total health expenditure per capita (USD)<br>Total public expenditure on education per capita (USD)<br>Healthy life expectancy<br>Total expenditure on R&D per capita (USD) | **O1 National appeal**<br>Parallel economy (IMD-WCI score)<br>Ethical practices (IMD-WCI score)<br>Investment risk (Euromoney score)<br>Image abroad (IMD-WCI score)<br>Country credit rating (Institutional Investor Magazine score)<br>Brain drain (IMD-WCI score)<br>Risk of political instability (IMD-WCI score) | **I1 Firm competitiveness**<br>Innovative capacity (IMD-WCI score)<br>Productivity of companies (IMD-WCI score)<br>Small and medium-size enterprises (IMD-WCI score)<br>Information technology skills (IMD-WCI score)<br>Large corporations (IMD-WCI score) |
| **F2 Quality of education system** | **O2 Investment environment** | **I2 Government intervention**<br>Subsidies (IMD-WCI score) |
| Educational assessment / Mathematics (PISA, 15-year-olds)<br>Educational assessment / Sciences (PISA, 15-year-olds)<br>Science in schools (IMD-WCI score)<br>Educational system (IMD-WCI score) | Foreign investors (IMD-WCI score)<br>Exchange rate stability (National currency/SDR parity change)<br>Capital markets (IMD-WCI score)<br>Investment incentives (IMD-WCI score)<br>State ownership of enterprises (IMD-WCI score) | Finance and banking regulation (IMD-WCI score)<br>Protectionism (IMD-WCI score)<br>Legal and regulatory framework (IMD-WCI score)<br>Ease of doing business (WB score)<br>Bureaucracy (IMD-WCI score) |
| **F3 Responsible enterprises**<br>Ethical practices (IMD-WCI score)<br>Social responsibility (IMD-WCI score)<br>Credibility of managers (IMD-WCI score) | | **I3 Availability of resources**<br>Labour force (percent of population)<br>Total primary energy supply per capita (toe per capita)<br>Burden of government regulation (IMD-WCI score)<br>Employment rate (percent of working age population)<br>Gross domestic savings (percent of GDP) |
| Extraction Method: Principal Component Analysis<br>Rotation Method: Varimax with Kaiser Normalisation | | |
| KMO: 0.823<br>Total variance explained: 83.9% | KMO: 0.793<br>Total variance explained: 73.7% | KMO: 0.791<br>Total variance explained: 73.4% |

Source: own work

The factor structure is estimated once, on the 2010 cross-section. 2010 is the natural choice because it is the year in which the intermediate clusters first separate from the two extremes, so it is where the transition strategies become visible and where the factors that distinguish them can be identified. The resulting factors are then used to characterise all clusters; they are a fixed interpretive lens, not a set re-estimated for each period. The analysis rests on 115 indicators (Appendix 1), each correlated with at least one FOI index, entered into an exploratory factor analysis.

The future potential of an OECD economy is influenced by three factors: F1 Human capital potential, the F2 Quality of the education system, and the F3 Socially responsible practices of enterprises. The outside potential is largely affected by the O1 National appeal (aggregating the impact of factors that only change over the long term), and the O2 Investment environment (which can change more quickly). Finally, the inside potential is shaped by I1 Firm competitiveness, I2 Government intervention, and I3 Availability of resources. The Kaiser-Meyer-Olkin (KMO) index for the three factor analyses was approximately 0.8 (ranging from 0.791 to 0.823), indicating the data is well-suited for factor analysis. The identified factors explained over 73% of the total variance in the FOI indices (ranging from 73.4% to 83.9%), which is also considered to be a good outcome. Appendix 2 shows the factor loadings of the rotated component matrices.

With the factor structure determined, standardised component scores were generated for all countries by the regression method (Thomson scoring). The resultant component scores serve as indicators of a country's performance relative to the mean, with zero signifying average performance, positive scores denoting above-average performance, and negative scores reflecting below-average performance. Cluster averages were then calculated from the country-level values within each cluster. These averages are presented in Table 6 (averages are reported for the three clusters central to the CEE transition plus FOI, the top cluster).

Table 6. Factor averages calculated for three OECD clusters

| Cluster | F1 | F2 | F3 | O1 | O2 | I1 | I2 | I3 |
|---|---|---|---|---|---|---|---|---|
| FOI | 0.57 | 0.27 | 0.88 | 1.03 | 0.23 | 0.55 | 0.70 | 0.58 |
| FOi | 0.28 | 0.50 | -0.12 | 0.17 | 0.27 | 0.64 | -0.51 | -0.40 |
| fOi | -0.71 | -0.11 | -1.10 | -0.99 | 0.10 | -1.00 | -0.49 | -0.13 |
| foi | -0.55 | -1.60 | -0.08 | -1.24 | -0.37 | -0.84 | -0.43 | -1.20 |

Source: own calculations

The outside-oriented strategy represented by the fOi cluster demonstrates above-average performance only in O2 Investment environment. This highlights the strategic importance of creating favourable conditions for foreign investors – the externally financed, infusion-stage configuration that the trap literature identifies as viable for entering middle income but insufficient for leaving it (Table 1, Gill–Kharas and 3i rows). Although the cluster shows moderate advantages over the lowest-performing group in F2 Quality of education system, and I3 Availability of resources, both factors fall short of overall averages. This indicates that these countries possess latent capacity for further economic transformation through technology adoption, supported by an education system that, while slightly below the OECD average, offers a comparative advantage.

This possible advantage offered by the education system is exploited very efficiently by the countries of the FOi cluster. Their strategy is a combination of external orientation, robust government intervention, and cutting-edge education system. The external orientation is an element taken from the previous stage of development. The other two components of this strategy are based on the government’s active role in social and economic management. The presence of an efficient developmental state, providing selective advantages to domestic actors (Kohli, 2009; Larson & Park, 2014; Wade, 2018), underscores this role. The success of this approach, notably in South Korea (Noland, 2012), has renewed interest in industrial policy (Juhász et al., 2023; Shih, 2023).

The core of this transition strategy is the successful implementation of industrial policy. The state's more prominent role is reflected in a low I2 Government intervention score, indicating above-average government involvement in subsidies, market protection, and regulation. In the terms of Table 1, a low I2 score marks the developmental state correction of the incentive distortions and talent misallocation Agénor (2017) identifies as a trap mechanism: active industrial policy steering resources and skills toward innovation rather than leaving them in protected or rent-seeking sectors. The fact that only South Korea and Estonia achieved successful transition to the FOi cluster shows the difficulty of implementing industrial policy in a manner that preserves external competitiveness.

In addition to implementing effective industrial policy, the developmental state prioritizes the creation of a competitive education system. This is reflected in the FOi cluster's superior F2 Quality of education system factor score, which is the highest among OECD nations within this group. Although the F1 Human capital factor is only slightly above average in the FOi cluster, the strong education system offers potential for further improvement. From an MIT perspective, human capital is paramount, as the recommended strategies for addressing its limitations – namely, increased TFP (Lee & Park, 2024), increased innovation activity (World Bank, 2024c), and higher retention of domestic value added (Gao et al., 2024) – are all linked to it. This is the human capital quality constraint Agénor (2017) places at the centre of the innovation transition (Table 1): not the quantity of schooling but its quality, which the F2 factor captures directly.

Education investment supports inclusive growth and the development of inclusive institutions (Serneels & Dercon, 2020; Shi & Qamruzzaman, 2022). Increased educational investment may therefore also contribute to overcoming the MIT via institutional channels (Mickiewicz, 2023; Song et al., 2023), specifically by reducing the ability of incumbents to impede economic transition. Education can also help bridge the skills gap, a fundamental cause of increasing inequality within the Kuznets curve model of economic transition (Kuznets, 1955).

### 4.3. Configurations and strategies in the CEE members of the OECD

The PCA of Section 4.2 identifies two broad development strategies among the OECD economies, and the CEE members sit at the divide between them: the outside-oriented strategy of the fOi cluster, which scores above the OECD average on the O2 investment-environment factor, and the innovation-oriented strategy of the FOi cluster, whose members recorded the highest F2 quality of education scores in the OECD in 2010. The first is viable for entering middle income but not for leaving it; the move from one to the other is the investment-to-innovation switch made observable as a cluster transition.

In 2000 five of the six CEE economies occupied the all-low foi cluster; Slovenia was the exception, holding high future potential alone (Foi). Over the following two decades the region turned decisively outward, raising its outside potential through deeper trade and financial integration. By 2020 five of the six had crossed the outside threshold: Czechia, Hungary, Slovakia and Slovenia hold the outside-only fOi configuration, and Estonia added high future potential to reach FOi. The sixth, Poland, sits fractionally below it.

Poland reached the fOi cluster in 2010 but by 2020 its outside index had eased to 3.9995, returning it, under the strict rule, to the foi cluster. A Monte-Carlo perturbation of its components (Section 4.5) splits it almost evenly between the two, so Poland is best read as sitting on the fOi boundary alongside its regional peers, not as having fallen back from them; its relative income in fact rose by 24 percentage points of the US level over the period, the second-strongest gain in the group. Slovenia presents the mirror boundary on the future dimension: its 2020 future index of 3.99 leaves it just inside the fOi cluster, but the same perturbation gives it a 39 per cent probability of reaching the future-high FOi cluster – a level it last held in 2000, before its future potential slipped and then partially recovered. Both are therefore treated as border cases rather than assigned to a single cluster. Estonia, by contrast, requires no such qualification: it is the only CEE economy to complete the transition, and it recorded the region's largest income convergence (a 27-point gain).

Table 7. Indicators of the Future and Outside potential of the CEE countries

| Country | F2_2010 | O2_2010 | F_2020 | O_2020 | ΔF (2000 – 2020) | ΔO (2000 – 2020) |
|---|---|---|---|---|---|---|
| Czechia | -0.1 | 0.6 | 3.8 | 4.2 | 0.7 | 1.8 |
| Estonia | 0.5 | 0.1 | 4.2 | 4.7 | 1.1 | 1 |
| Hungary | -0.7 | -0.1 | 3.1 | 4.4 | -0.3 | 1.2 |
| Poland | 0.0 | 0.0 | 3.7 | 4 | 0.5 | 0.8 |
| Slovakia | -0.9 | n.a. | 3.4 | 4.8 | -0.2 | 2.2 |
| Slovenia | -0.3 | -1.7 | 4 | 4.5 | -0.1 | 1.3 |

Source: own calculation

Table 7 traces the paths in detail. Outside potential rose across all six economies, but progress on the future dimension divided them. Estonia's education-quality factor was well above the OECD average in 2010; Czechia and Poland were near it and improving; Hungary, Slovakia and Slovenia scored below average on future potential, though Slovenia's position is a boundary case rather than a settled low. This internal divide is not an artefact of the FOI method: three further studies, using entirely different approaches, identify the same grouping. Győrffy (2022), analysing the region's institutions qualitatively, names Czechia, Slovenia, Estonia and Lithuania as the economies that stepped onto a productivity- and innovation-led path in the 2010s, while others relied on low costs, EU transfers and loose monetary conditions. A ranking of CEE-EU innovation potential by a multi-criteria (TOPSIS) method over 2013–2022 places Slovenia, Czechia and Estonia in the top innovation group, investing in R&D and social development, with the remainder lagging (Brodny & Tutak, 2024). And an analysis of Visegrad industrial policy finds autonomous research-and-innovation capability emerging where policy has turned to RDI – clearest in Czechia, with Poland the most domestically focused – and remaining marginal where it has not, in Hungary and Slovakia (Szabó et al., 2025). The fact that studies using four different methods independently recover the same escaper–laggard divide, is strong evidence that the FOI configurations are tracking a real feature of the region's development.

The mechanism behind the divide is the one the framework predicts. External integration without the parallel build-up of future potential is the fOi configuration, and a growing literature documents its limits in the region (Győrffy, 2022; Szabó et al., 2025). This is the Lewis-type erosion of a cost-based external configuration made concrete, and it is why the outside turn alone does not secure escape.

The two CEE economies the framework places closest to escape illustrate the F-dimension mechanism directly. Estonia's transition rests on a sustained investment in human capital and innovation capacity: living standards have doubled since 2000, R&D expenditure reached 1.8 per cent of GDP, and the education system has been repeatedly reoriented toward the skills that frontier growth requires – a strategy whose continuity into the 2020s is visible in the vocational and digital-skills reforms tied to Industry 4.0 (European Commission, 2025; OECD, 2024). Estonia implemented institutional reforms in its education system during the 1990s which led to the establishment of an external evaluation system, designed to provide feedback on students' achievements (Tire, 2020), and granted high autonomy to school leaders in setting strategies for school improvements (Vanari & Eisenschmidt, 2024). It was found that these reforms resulted in a 15-30% improvement in adult literacy, numeracy, and problem-solving skills (Byrne & Plekhanov, 2019). Estonia now performs strongly in digital literacy, adult participation in learning, and in minimising academic underachievement (Streimikiene & Bathaei, 2025). This is the F2 education-quality factor of the 2010 analysis persisting as a development strategy.

Slovenia's cusp position has a firm empirical basis in the future-oriented investments the F-dimension measures. Its gross domestic expenditure on research and development reached 2.1 per cent of GDP in 2023 – close to the EU average of 2.24 per cent and by a wide margin the highest R&D intensity in Central and Eastern Europe – with roughly two-thirds of that spending performed by the business-enterprise sector (Eurostat, 2025; SURS, 2024). This build-up of innovation capacity has been accompanied by sustained productivity convergence: Slovenia's labour-productivity gap to the OECD average narrowed by thirteen percentage points between 2010 and 2023, with within-industry productivity growth outpacing the OECD benchmark over the period (OECD, 2026). These are the investments that place Slovenia on the boundary of the future-high group rather than within the merely outside-oriented one, and they are why the qualitative and multi-criteria literatures alike assign it to the innovation-led path even though the framework's fixed threshold, read strictly, places its 2020 future index (3.99) just below the line.

### 4.4 Validation: configurations, income levels and convergence

The FOI configurations are defined without reference to income, and the argument of Section 2.1 was that they complement income-based trap identification rather than reproduce it. That argument carries an empirical obligation: if configurations are to speak to the middle-income trap, they must at least co-move with income levels and with convergence in the directions the mapping predicts. This section examines that co-movement. It is a consistency check, not a causal test – with thirty-four economies and three observation points, the aim is to establish that the configuration space is not detached from the income dynamics the trap literature is concerned with.

Each country's income is expressed as its GDP per capita at purchasing-power parity relative to the United States, the frontier economy, in the same year. Relative income isolates convergence from common global growth: a country can grow in absolute terms yet lose ground if the frontier advances faster, and it is this relative movement that the trap concerns. Each country's configuration is summarised by its configuration score – the number of its three potentials that are high (0 to 3) – which orders the eight clusters by structural development without imposing finer assumptions.

Table 8 reports the median relative income of each configuration score at each date. The ordering is monotonic in every year: higher configurations correspond to higher relative income, and the gap between the lowest and highest configurations is large and stable – roughly forty percentage points of US income separate score-0 from score-3 economies throughout. The rank correlation between configuration score and relative income is strong and highly significant at each date (Spearman $\rho = 0.78$, 0.75 and 0.73 for 2000, 2010 and 2020; $p < 0.001$). Structural configuration and income position are thus tightly aligned, as the mapping in Section 2.1 requires.

Table 8. Median relative income (% of US GDP-PPP per capita) by configuration score

| Configuration score | 2000 | 2010 | 2020 |
|---|---|---|---|
| 0 (all potentials low) | 39% (n=11) | 50% (n=4) | 50% (n=6) |
| 1 (one high) | 61% (n=4) | 54% (n=8) | 61% (n=7) |
| 2 (two high) | — | 78% (n=11) | 77% (n=9) |
| 3 (all potentials high) | 87% (n=19) | 94% (n=11) | 92% (n=12) |

*Notes: median GDP per capita at PPP as a percentage of the US level. No economy held a two-high configuration in 2000. Spearman ρ (score, relative income) = 0.78 / 0.75 / 0.73; $p < 0.001$.*

Source:

Source: own calculations

The dynamic counterpart is whether movement between configurations accompanies convergence. Grouping the thirty-four economies by their configuration trajectory over 2000–2020 (Table 9) produces the expected ordering. The seven upward movers – economies that added at least one high potential, and which include four of the six CEE members – converged most strongly, gaining a median of eighteen percentage points of US income. The eighteen persistent-high economies, already at the frontier, were essentially flat (median +0.6 points), as convergence leaves little room at the top. The seven persistent-low economies diverged, losing a median of six points. The pattern is that configurational upgrading and income convergence travel together, and its absence and income stagnation travel together.

Table 9. Configuration trajectory and income convergence, 2000–2020

| Trajectory group | n | Median Δ relative income (pp) | of which CEE |
|---|---|---|---|
| Upward mover (score rose) | 7 | +18.3 | Czechia, Estonia, Hungary, Slovakia |
| Persistent-high (score ≥2 throughout) | 18 | +0.6 | — |
| Downward mover (score fell) | 2 | +4.5 | — |
| Persistent-low (score ≤1 throughout) | 7 | −5.7 | Poland, Slovenia |

*Notes: Δ relative income is the change in GDP-PPP per capita as a share of the US level, in percentage points, 2000–2020. Only the between-group ordering is interpreted; the rank correlation between the change in configuration score and the change in relative income is positive but not significant (ρ = 0.31, p = 0.08), the ordinal score being too coarse to register within-cluster convergence such as Poland's.*

Source: own calculations

Finally, the framework's specific trap diagnoses hold in income terms. The three economies that remained in the all-low foi configuration throughout – Greece, Italy and Turkey – are instructive because they diverge. Greece and Italy are the textbook cases: high-income economies whose relative position eroded sharply (Italy from 87 to 69 per cent of US income, Greece from 57 to 46) while their structural configuration never improved – the configurational signature of the trap, corroborated by income decline. Turkey is the exception. It converged by fifteen points, halving its distance to the frontier, yet without any structural upgrading: income growth without configurational change. Income-based identification alone would read this as escape; the configuration lens reads it as growth that the Schumpeterian and institutional accounts predict is difficult to sustain, because the capabilities that carry an economy to the frontier were not built. The FOI model is well suited to reveal this kind of divergence between income status and structural capability.

Three limitations bound these findings. The sample is thirty-four economies observed at three points, so the analysis is descriptive rather than inferential. The co-movement does not identify causation: capability building may raise income, higher income may fund capability building, or both may respond to a common cause. And two income observations warrant caution: Ireland's post-2015 figures are inflated by the relocation of multinational intellectual property, and Luxembourg's by cross-border labour – though as frontier economies neither affects the configuration–income ordering that the analysis turns on. Within these bounds, the configurations behave as the trap mapping requires: they track income levels, upgrading accompanies convergence, and structural persistence accompanies the trap.

### 4.5 Robustness of the classification

The interval-halving classification rests on two choices: the fixed threshold of 4 that separates high from low, and the min-max standardisation, whose endpoints are set by the most extreme country on each indicator. The sensitivity of the classification was tested to both, and to two further sources of uncertainty, across all three observation years. The tests are summarised in Table 10 and reported in full in Appendix 3. The conclusion is that the cluster assignments are stable, with movements confined to countries already at the threshold.

First, because the fast-moving components of the Inside and Outside indices (real growth, government revenue, trade openness) could be distorted by shocks around the specific year sampled, those components were recomputed as five-year trailing averages. This changed three of thirty-two assignments in 2020 and three in 2010, every one of them a country whose Inside index already lay within half a point of the threshold, and none of them a CEE member. The Future index, which contains no fast-moving component, was unaffected.

Second, the threshold itself was varied. Twenty of the 306 country-year-index values lie within 0.1 of four, and fifty-six within 0.25. Shifting the threshold by a quarter-point reassigns between a fifth and a third of cases, but exclusively among these borderline countries; the cores of the clusters do not move.

Third, the influence of the extreme countries that set the standardisation endpoints was tested with two methods: winsorising each indicator at its fifth and ninety-fifth percentiles (OECD & European Commission JRC, 2008) and separately recomputing the anchors with outlier countries removed (one at a test). Winsorising changed a single assignment in 2010 and one in 2020 – Luxembourg, whose extreme values are the reason for the procedure – and none in 2000. The leave-one-out exercise moved between two and six borderline countries depending on which extreme country was withheld, confirming that endpoint sensitivity, a known property of min-max standardisation, is real but confined to the boundary.

Finally, the assignment uncertainty of the borderline countries was calculated directly. Every component was perturbed with a random measurement error (Gaussian, scaled to five per cent of the component's cross-country standard deviation) and countries were reclassified based on ten thousand draws (Saisana et al., 2005). Fifteen of the nineteen countries in the 2020 corridor retained their assignment in every draw. The four exceptions are the genuine boundary cases: Canada (50 per cent FOI, 50 per cent fOI), Poland (51 per cent foi, 49 per cent fOi), Slovenia (61 per cent fOi, 39 per cent FOi), and Spain (72 per cent fOi, 28 per cent foi).

As a cross-check, the interval-halving partition was compared with k-means and Ward clustering of the same indices. The data-driven methods prefer fewer, broader clusters than the interval-halving grid. Where the structure is a clean high–low split, the two approaches agree strongly (in 2000 the Adjusted Rand Index reaches 0.80). In 2010 and 2020, where the interval-halving distinguishes six configurations but the data support only two or three natural groups, agreement is moderate (ARI 0.39–0.59), and the disagreements fall almost entirely on the intermediate clusters.

Taken together, the tests show that the classification is robust to the temporal sampling of its cyclical components, to the position of the threshold, and to the extreme countries that anchor the standardisation. The sensitivity that exists is not diffuse: it is concentrated on a small, identifiable set of boundary countries, which the analysis treats explicitly as such.

Table 10. Summary of robustness tests (all three years)

| Test | What it varies | Effect on classification |
|---|---|---|
| Five-year averaging | Cyclical components replaced by 5-year means | 3 reassignments in 2010 and in 2020, all at the I-threshold; no CEE country; F unaffected |
| Threshold sensitivity | High/low cut-off shifted ±0.25 | 56 of 306 index values lie within 0.25 of the cut-off; reassignments confined to borderline countries |
| Winsorised standardisation | Indicators clipped at 5th/95th pct | 1 reassignment in 2010 and in 2020 (Luxembourg); 0 in 2000 |
| Leave-one-out anchoring | Standardisation endpoints recomputed excluding outlier countries one-by-one | 2–6 borderline reassignments depending on the excluded country |
| Monte-Carlo perturbation | ±5% measurement error, 10,000 draws | 15 of 19 corridor countries stable in every draw; 4 boundary cases quantified |

*Notes: full results in Appendix 3. "Corridor" countries are those whose decisive index lies within 0.25 of the threshold.*

## 5. Conclusion

Emerging economies frequently experience a persistent development slowdown, a phenomenon known as the middle-income trap. It is usually identified by comparing a country's economic output to an absolute or relative income threshold. Income thresholds, however, reveal that convergence has stalled without indicating where in an economy's structure the blockage lies. This study takes a different route, treating the trap as a configuration problem: the six major theoretical accounts of the middle-income trap each identify a growth configuration that is viable for entering middle income, a different configuration required for leaving it, and locate the trap in the failure to move between them (Table 1). The FOI model, which resolves development into future, outside, and inside potential, provides a measurement surface on which those configurations become observable and comparable. The model itself is established (Bartha, 2024; Bartha & S. Gubik, 2014); what this study adds is the creation of a three-period panel in which configuration change can be observed; an explicit mapping between the trap literature's mechanisms and the model's dimensions; the application of both to the middle-income trap question in the Central and Eastern European members of the OECD; and a validation of the resulting configurations against income convergence. Using statistics from 34 OECD countries over 2000–2020, two potential traps and two transition strategies were identified.

The first trap affects economies with low future, outside, and inside potential – the foi cluster. These countries lack both the internal resources and the domestic demand to drive growth, and persistence in this configuration across the whole period is the configurational signature of the trap, translating the duration criterion of income-based definitions into structural terms. A strategy focused on external resources offers a way out. Countries with high outside potential, forming the fOi cluster, demonstrate this approach: an attractive investment environment draws foreign capital and technology, which are then used to boost exports. This is the infusion-stage configuration of the trap literature (Table 1, Gill–Kharas and 3i rows). By 2010 all six CEE OECD members – Czechia, Estonia, Hungary, Poland, Slovakia and Slovenia – had adopted it.

The outside-focused strategy can itself become a trap, the second identified here. Excessive reliance on external resources makes it increasingly difficult to raise domestic value added, while rising wages, driven by the Lewis-effect, erode export competitiveness. The configuration is transitory by construction (Table 1, Lewis and dependency rows): the advantages that make it viable are the same advantages that its own success destroys. Four of the six CEE economies remained in this configuration in 2020.

Escaping it requires the second transition strategy, represented by the FOi cluster. This builds on the external orientation of the fOi countries and integrates it with a developmental-state approach centred on the quality of the education system – the investment-to-innovation switch that the Schumpeterian account places at the core of the escape, and the human-capital-quality constraint that Agénor (2017) identifies as binding. Better human capital raises total factor productivity, improves innovation efficiency, and allows an economy to retain a higher share of domestic value added. Implementing it demands greater government involvement, which can in turn depress the inside potential – a tension that plausibly explains why so few economies complete the move. Estonia is the only CEE country to have done so, reaching the FOi cluster by 2020 and recording the region's largest income convergence, a gain of 27 percentage points relative to the United States. Czechia and Poland show positive movement on the future dimension without yet crossing the threshold, and Slovenia sits on the boundary of the future-high group, a position corroborated by the qualitative, multi-criteria and industrial-policy literatures alike, which independently assign the same set of economies to the innovation-led path.

The configurations are not detached from the income dynamics the trap literature is concerned with. Median relative income rises monotonically with the number of high potentials at all three observation points, and economies that added a high potential over the period converged most strongly (a median gain of 18 percentage points of US income), while those that remained in a low configuration diverged (a median loss of six points). The framework's specific diagnoses also hold: Greece and Italy, which never left the all-low configuration, lost ground sharply against the frontier, while Turkey converged without any structural upgrading – growth that income-based identification alone would read as escape, and that the configuration lens reads as fragile.

The method has three principal limitations. The analysis rests on statistical association and cannot establish causality: capability building may raise income, higher income may fund capability building, or both may respond to a common cause. Second, reliance on indicators whose definitions, reliability and availability evolve complicates long-term comparison. Third, the relative construction of the indices — anchored by the best and worst performers in the reference group — introduces sensitivity to the group's composition and to the position of the high–low threshold. This third limitation has been examined directly: the classification proves robust to five-year averaging of its cyclical components, to winsorising and leave-one-out re-anchoring of the standardisation, and to simulated measurement error, with the sensitivity that exists confined to a small, identifiable set of boundary countries rather than diffused through the sample (Section 4.5).

Within these bounds, the FOI model makes a distinct contribution to the middle-income-trap debate in the CEE region. By measuring development along three dimensions rather than one, it tracks the structural shifts that income figures cannot see, distinguishes convergence from capability building where the two diverge, and identifies the transition strategies that have actually worked. Its practical implication is direct: for the CEE economies still holding the outside-oriented configuration, the external opening that carried them this far will not carry them further, and the binding constraint lies in the quality of human capital and the institutions that allocate it.

Funding declaration

Project no. K 147241 has been implemented with the support provided by the Ministry of Culture and Innovation of Hungary from the National Research, Development and Innovation Fund, financed under the „OTKA" (K_23) funding scheme.

## Appendix 1 — Variables included in the exploratory factor analysis

The 115 indicators screened for the factor analysis are listed below. Source abbreviations are expanded in the key at the end of this appendix.

| Variable | Source | Description |
|---|---|---|
| Access to commodities | IMD-WCY | Access to commodities meets the needs of business (IMD executive survey, 0–10 scale). |
| Adaptability of companies | IMD-WCY | Adaptability of companies to market changes is high (IMD executive survey, 0–10 scale). |
| Ageing of society | IMD-WCY | Ageing of society is not seen as a burden for economic development (IMD executive survey, 0–10 scale). |
| Attitudes toward globalisation | IMD-WCY | Attitudes toward globalisation are generally positive in society (IMD executive survey, 0–10 scale). |
| Brain drain | IMD-WCY | Emigration of skilled people does not hinder competitiveness (IMD executive survey, 0–10 scale). |
| Bribing and corruption | IMD-WCY | Bribing and corruption do not exist (IMD executive survey, 0–10 scale). |
| Burden of government regulation | IMD-WCY | Compliance with government administrative requirements is not burdensome for business (survey, 1–7 scale). |
| Bureaucracy | IMD-WCY | Bureaucracy does not hinder business activity (IMD executive survey, 0–10 scale). |
| Capital markets | IMD-WCY | Foreign and domestic capital markets are easily accessible (IMD executive survey, 0–10 scale). |
| Central government foreign debt | IMD-WCY | Central government foreign debt, % of GDP. |
| CO2 emissions intensity | IMD-WCY | CO2 emissions per unit of GDP. |
| Collected total tax revenues | OECD | Total tax revenue, % of GDP. |
| Communications technology | IMD-WCY | Communications technology meets business requirements (IMD executive survey, 0–10 scale). |
| Compensation levels | IMD-WCY | Compensation levels in industry are adequate by international comparison (IMD executive survey, 0–10 scale). |
| Competition legislation | IMD-WCY | Competition legislation is efficient in preventing unfair competition (IMD executive survey, 0–10 scale). |
| Corporate tax rate on profit | IMD-WCY | Statutory corporate tax rate on profits, %. |
| Country credit rating | IIM | Country creditworthiness rating, 0–100 scale (Institutional Investor). |
| Credibility of managers | IMD-WCY | Credibility of managers in society is high (IMD executive survey, 0–10 scale). |
| Current account balance | IMF | Current account balance, % of GDP. |
| Customs authorities | IMD-WCY | Customs authorities do not hinder the efficient transit of goods (IMD executive survey, 0–10 scale). |
| Dependency ratio | IMD-WCY | Population aged under 15 and over 64, % of working-age population. |
| Direct investment stocks abroad | IMD-WCY | Outward FDI stock, % of GDP. |
| Direct investment stocks inward | IMD-WCY | Inward FDI stock, % of GDP. |
| Ease of access to loans | IMD-WCY | A good business plan without collateral is sufficient to obtain a bank loan (survey, 1–7 scale). |
| Ease of doing business | WB | Ease of doing business is supported by regulation (IMD executive survey, 0–10 scale). |
| Ecological footprint | GFN | Global hectares of biocapacity demanded per person. |
| Educational assessment — Mathematics | OECD (PISA) | PISA survey mean score, mathematics, 15-year-olds. |
| Educational assessment — Sciences | OECD (PISA) | PISA survey mean score, science, 15-year-olds. |
| Educational system | IMD-WCY | The educational system meets the needs of a competitive economy (IMD executive survey, 0–10 scale). |
| Employment in the public sector | IMD-WCY | Public-sector employment, % of total employment. |
| Employment rate | OECD | Employed population, % of working-age population. |
| Energy infrastructure | IMD-WCY | Energy infrastructure is adequate and efficient (IMD executive survey, 0–10 scale). |
| English proficiency (TOEFL) | ETS | Mean TOEFL score. |
| Entrepreneurship | IMD-WCY | Entrepreneurship among managers is widespread in business (IMD executive survey, 0–10 scale). |
| Ethical practices | IMD-WCY | Ethical practices are implemented in companies (IMD executive survey, 0–10 scale). |
| Exchange rate stability | IMF (IFS) | Change in parity between the national currency and the SDR (three-year average) |
| Exports of commercial services | IMD-WCY | Commercial services exports, US$ billions. |
| Exports of goods | IMD-WCY | Goods exports, US$ billions. |
| Finance and banking regulation | IMD-WCY | Finance and banking regulation facilitates business activity (IMD executive survey, 0–10 scale). |
| Financial institutions' transparency | IMD-WCY | Financial institutions' transparency is sufficiently implemented (IMD executive survey, 0–10 scale). |
| Flexibility and adaptability | IMD-WCY | Flexibility and adaptability of people are high when faced with new challenges (IMD executive survey, 0–10 scale). |
| Foreign currency reserves | IMF | Foreign exchange reserves. |
| Foreign high-skilled people | IMD-WCY | The economy attracts skilled foreign workers (IMD executive survey, 0–10 scale). |
| Foreign investors | IMD-WCY | Foreign investors are welcome (IMD executive survey, 0–10 scale). |
| Foreign labour force | IMD-WCY | Foreign labour force, % of total labour force. |
| Future energy supply | IMD-WCY | Future energy supply is secure (IMD executive survey, 0–10 scale). |
| GDP | IMF (WEO) | Gross domestic product, US$ billions. |
| GDP at PPP | IMF (WEO) | Gross domestic product at purchasing power parity, US$ billions. |
| General government gross debt | IMF | General government gross debt, % of GDP. |
| Gini index | IMD-WCY | Income inequality |
| Government subsidies | IMD-WCY | Subsidies to private and public companies, % of GDP. |
| Gross domestic savings | OECD | Gross domestic savings, % of GDP. |
| Healthy life expectancy | WHO | Average estimated years of life in full health. |
| Higher education achievement | IMD-WCY | Share of population with higher education attainment. |
| Human development index | UN | Composite of income, education and health indicators (UNDP Human Development Report). |
| Image abroad | IMD-WCY | The country's image abroad encourages business development (IMD executive survey, 0–10 scale). |
| Immigration laws | IMD-WCY | Immigration laws do not prevent the country from attracting skilled labour (IMD executive survey, 0–10 scale). |
| Industrial disputes | IMD-WCY | Working days lost to industrial disputes per 1,000 inhabitants |
| Information technology skills | IMD-WCY | Information technology skills are readily available (IMD executive survey, 0–10 scale). |
| Innovative capacity | IMD-WCY | Firms' capacity to generate new products, processes or services is high (IMD executive survey, 0–10 scale). |
| Investment incentives | IMD-WCY | Investment incentives are attractive to foreign investors (IMD executive survey, 0–10 scale). |
| Investment risk | IIM | Country creditworthiness scale, 0–100 (Euromoney). |
| Labour force | OECD (LFS) | Labour force, % of total population. |
| Labour productivity (PPP) | IMD-WCY / PPP | GDP at purchasing power parity per person employed, per hour worked (US$). |
| Labour regulations | IMD-WCY | Labour regulations do not hinder business activity (IMD executive survey, 0–10 scale). |
| Language skills | IMD-WCY | Language skills meet the needs of enterprises (IMD executive survey, 0–10 scale). |
| Large corporations | IMD-WCY | Large domestic companies are competitive in world markets (IMD executive survey, 0–10 scale). |
| Legal and regulatory framework | IMD-WCY | The legal and regulatory framework encourages competitiveness (IMD executive survey, 0–10 scale). |
| Life expectancy at birth | WHO | Average estimated years of life at birth. |
| Market value of publicly traded shares | IMD-WCY | Stock market capitalisation. |
| National culture | IMD-WCY | The national culture is open to foreign ideas (IMD executive survey, 0–10 scale). |
| Natural gas exports | EIA | Natural gas exports. |
| Number of patents in force | IMD-WCY | Patents in force per 100,000 inhabitants. |
| Office rent | IMD-WCY | Average annual office rent per square metre. |
| Overall productivity (PPP) | IMD-WCY / PPP | GDP at purchasing power parity per person employed (US$). |
| Parallel economy | IMD-WCY | The parallel (informal) economy is not a threat to the formal economy (IMD executive survey, 0–10 scale). |
| Patent applications | IMD-WCY | Number of patent applications filed by residents and non-residents. |
| Patent productivity | IMD-WCY | Patents granted to residents per thousand R&D personnel. |
| Pension funding | IMD-WCY | Pension funding adequately addresses future needs (IMD executive survey, 0–10 scale). |
| Productivity of companies | IMD-WCY | Companies' productivity is supported by global strategies — supply chains, offshoring, outsourcing (IMD executive survey, 0–10 scale). |
| Protectionism | IMD-WCY | Protectionism does not impair the conduct of business (IMD executive survey, 0–10 scale). |
| Public sector contracts | IMD-WCY | Public sector contracts are open to foreign bidders (IMD executive survey, 0–10 scale). |
| Pupil-teacher ratio (primary education) | IMD-WCY | Number of pupils per teacher, primary education. |
| Pupil-teacher ratio (secondary education) | IMD-WCY | Number of pupils per teacher, secondary education. |
| Quality of air transportation | IMD-WCY | Air transportation meets business needs (IMD executive survey, 0–10 scale). |
| Quality of life | IMD-WCY | Quality of life is high (IMD executive survey, 0–10 scale). |
| Real GDP growth | IMF (WEO) | Real GDP growth, % change, constant prices. |
| Real GDP per capita at PPP | IMF (WEO) | GDP per capita at purchasing power parity, US$. |
| Renewable energies | IMD-WCY | Share of renewables and waste in total energy requirements (%). |
| Reserves of foreign exchange and gold | WB | Foreign exchange and gold reserves. |
| Rigidity of employment | IMD-WCY | Employment regulation is not overly rigid |
| Risk of political instability | IMD-WCY | Political instability is not a risk to business activity (IMD executive survey, 0–10 scale). |
| Science in schools | IMD-WCY | Science is adequately taught in schools (IMD executive survey, 0–10 scale). |
| Skilled labour | IMD-WCY | Skilled labour is readily available (IMD executive survey, 0–10 scale). |
| Small and medium-size enterprises | IMD-WCY | SMEs are efficient by international standards (IMD executive survey, 0–10 scale). |
| Social responsibility | IMD-WCY | Social responsibility of business leaders is high (IMD executive survey, 0–10 scale). |
| Start-up days | WB | Number of days required to start a business |
| State ownership of enterprises | IMD-WCY | State ownership of enterprises does not hinder business activity (IMD executive survey, 0–10 scale). |
| Student mobility outbound | IMD-WCY | Outbound tertiary-level students as a share of the student population. |
| Subsidies | IMD-WCY | Subsidies to private and public companies as a share of GDP |
| Sustainable development | IMD-WCY | Sustainable development is a priority for firms (IMD executive survey, 0–10 scale). |
| Tariff barriers | IMD-WCY | Most-favoured-nation tariff, simple average rate. |
| Terms of trade index | IMD-WCY | Export prices relative to import prices. |
| Total expenditure on R&D per capita | OECD | R&D expenditure, US$ per capita. |
| Total health expenditure per capita | OECD | Health expenditure, US$ per capita. |
| Total primary energy supply per capita | IMD-WCY | Primary energy supply per capita. |
| Total public expenditure on education | OECD | Public education expenditure, % of GDP. |
| Total public expenditure on education per capita | OECD | Public education expenditure, US$ per capita. |
| Total R&D personnel per capita | IMD-WCY | R&D personnel (full-time equivalent) per 1,000 population. |
| Tourism receipts | IMD-WCY | International tourism receipts. |
| Trade-to-GDP ratio | WB | (Exports + imports) / (2 × GDP). |
| Value system | IMD-WCY | Shared values facilitate business relationships (IMD executive survey, 0–10 scale). |
| Water consumption intensity | IMD-WCY | Water consumption per unit of GDP. |
| Water transportation | IMD-WCY | Water transportation infrastructure meets business needs (IMD executive survey, 0–10 scale). |
| Working hours | IMD-WCY | Average annual working hours per employee. |

## Source abbreviation key

| Abbreviation | Full name |
|---|---|
| IMD-WCY | IMD World Competitiveness Yearbook |
| OECD | OECD databases (Education at a Glance, Revenue Statistics, Labour Force Statistics, PISA) |
| IMF | International Monetary Fund (World Economic Outlook; International Financial Statistics) |
| WB | World Bank (World Development Indicators; Doing Business) |
| WEF-GCI | World Economic Forum, Global Competitiveness Index |
| WHO | World Health Organization |
| GFN | Global Footprint Network |
| IIM | Institutional Investor magazine |
| ETS | Educational Testing Service |
| UN | United Nations |
| EIA | U.S. Energy Information Administration |
| PPP | Purchasing power parity calculations (various sources) |

Source: own work

## Appendix 2 — Factor loadings of the rotated component matrices

*Extraction: Principal Component Analysis. Rotation: Varimax with Kaiser normalisation. Loadings below 0.50 suppressed. Factor labels are interpretive.*

### F-index (Future potential)

| Factor | Variable | Loading |
|---|---|---|
| *F1: Human capital & productivity* | Total health expenditure per capita | 0.907 |
| | Overall productivity (PPP) | 0.891 |
| | Labour productivity (PPP) | 0.891 |
| | Total public expenditure on education p.c. | 0.849 |
| | Healthy life expectancy | 0.712 |
| | Total expenditure on R&D per capita | 0.707 |
| *F2: Quality of education system* | Educational assessment — Mathematics | 0.945 |
| | Educational assessment — Sciences | 0.932 |
| | Science in schools | 0.711 |
| | Educational system | 0.657 |
| *F3: Corporate responsibility* | Credibility of managers | 0.919 |
| | Social responsibility | 0.824 |
| | Ethical practices | 0.742 |

### O-index (Outside potential)

| Factor | Variable | Loading |
|---|---|---|
| *O1: Country reputation & stability* | Parallel economy | 0.910 |
| | Ethical practices | 0.888 |
| | Investment risk | 0.803 |
| | Image abroad | 0.779 |
| | Country credit rating | 0.770 |
| | Brain drain | 0.756 |
| | Risk of political instability | 0.750 |
| *O2: Investment environment* | Foreign investors | 0.886 |
| | Exchange rate stability | −0.805 |
| | Capital markets | 0.798 |
| | Investment incentives | 0.786 |
| | State ownership of enterprises | 0.719 |

### I-index (Inside potential)

| Factor | Variable | Loading |
|---|---|---|
| *I1: Business sophistication* | Innovative capacity | 0.869 |
| | Productivity of companies | 0.827 |
| | Information technology skills | 0.815 |
| | Small and medium-size enterprises | 0.710 |
| | Large corporations | 0.559 |
| *I2: Government intervention* | Subsidies | 0.802 |
| | Protectionism | 0.754 |
| | Finance and banking regulation | 0.729 |
| | Legal and regulatory framework | 0.707 |
| | Ease of doing business | 0.660 |
| | Bureaucracy | 0.636 |
| *I3: Labour & resources* | Labour force | 0.878 |
| | Total primary energy supply per capita | 0.716 |
| | Burden of government regulation | 0.685 |
| | Employment rate | 0.677 |
| | Gross domestic savings | 0.542 |

Source: Own calculations

## Appendix 3 - Robustness of the FOI classification

This appendix gives the country-level detail behind the robustness summary in Section 4.5: the specific reassignments under each test, the countries whose assignment is genuinely uncertain, and the full comparison with data-driven clustering.

### A3.1 Averaging of cyclical components

The Inside and Outside indices contain fast-moving components (real GDP growth, government revenue, trade openness) that may reflect the specific year sampled rather than a country's structural position; the concern is acute for 2020, a pandemic year. Recomputing these components as five-year trailing averages (2016–2020 and 2006–2010) and reclassifying gives the changes in Table A3-1.

Table A3-1. Cluster changes under five-year averaging

| Year | Country | Baseline | Averaged | Dimension crossing 4 |
|---|---|---|---|---|
| 2010 | Israel | fOi | fOI | I |
| 2010 | Norway | FOI | FOi | I |
| 2010 | Sweden | FOi | FOI | I |
| 2020 | Austria | FOi | FOI | I |
| 2020 | Belgium | fOi | fOI | I |
| 2020 | France | FOi | FOI | I |

Source: own calculations

### A3.2 Leave-one-country-out anchoring

Table A3-2 reports, for the five countries that most often set an endpoint in each year, the number of other countries whose assignment changes. The effect is modest and localised: excluding an extreme country moves between zero and six borderline countries, with the largest single effect (Iceland, 2010) reflecting its outlying low Outside score.

Table A3-2. Reassignments under leave-one-country-out anchoring (most influential countries)

| Year | Country excluded | Countries reassigned | Max index shift |
|---|---|---|---|
| 2000 | Luxembourg | 3 | 0.55 |
| 2000 | Mexico | 1 | 0.37 |
| 2000 | Japan | 1 | 0.37 |
| 2010 | Iceland | 4 | 0.94 |
| 2010 | Luxembourg | 2 | 0.63 |
| 2020 | Japan | 5 | 0.65 |
| 2020 | Greece | 5 | 0.61 |
| 2020 | Mexico | 4 | 0.53 |
| 2020 | Turkey | 4 | 0.54 |
| 2020 | Luxembourg | 3 | 0.35 |

Source: own calculations

A3.3 Monte-Carlo perturbation

Following the uncertainty-analysis approach to composite indicators (Saisana et al., 2005), each component was perturbed with independent Gaussian noise scaled to 5 per cent of that component's cross-country standard deviation, and countries were reclassified over 10,000 draws. The exercise targets the countries in the corridor, since interior countries never change. Table A3-3 reports every country whose assignment was uncertain in more than one draw in five (that is, a probability below 80 per cent for its baseline cluster). All other corridor countries retained their assignment in every one of the 10,000 draws.

Table A3-3. Countries with uncertain assignment under Monte-Carlo perturbation

| Year | Country | Cluster 1 (prob.) | Cluster 2 (prob.) |
|---|---|---|---|
| 2000 | France | FoI (77%) | FOI (23%) |
| 2010 | Finland | FOI (78%) | FOi (22%) |
| 2010 | Japan | FoI (76%) | Foi (25%) |
| 2010 | Mexico | foi (71%) | fOi (29%) |
| 2010 | New Zealand | FOI (56%) | FOi (44%) |
| 2020 | Canada | FOI (50%) | fOI (50%) |
| 2020 | Poland | foi (51%) | fOi (49%) |
| 2020 | Slovenia | fOi (61%) | FOi (39%) |
| 2020 | Spain | fOi (72%) | foi (28%) |

Source: own calculations

A3.4 Comparison with data-driven clustering

As a final check, the theory-driven interval-halving partition was compared with k-means and Ward hierarchical clustering of the same three indices, using the Adjusted Rand Index (ARI) to measure agreement. The comparison is reported in Table A3-4. The data-driven methods favour a coarser partition than the interval-halving grid: by the silhouette criterion the preferred number of clusters is two or three, corresponding to the dominant high–low divide.

Table A3-4. Agreement between interval-halving and data-driven clustering (ARI)

| Year | ARI, k = 2 | Best ARI (k) | Interval-halving clusters |
|---|---|---|---|
| 2000 | 0.80 | 0.80 (k = 2) | 5 |
| 2010 | 0.36 | 0.59 (k = 8) | 6 |
| 2020 | 0.41 | 0.53 (k = 5) | 6 |

*Notes: ARI = Adjusted Rand Index (1 = identical partitions, 0 = chance agreement). k-means with 50 restarts; Ward linkage on Euclidean distance. Best ARI is the maximum over k = 2–8 across both methods.*

Source: own calculations